\documentclass[12pt]{article}

\usepackage{amsmath}
\usepackage{amssymb}

\begin{document}

\title{Thermodynamic inequalities for nuclear rotation
\footnote{Zh. Eksp. Teor. Fiz. {\bf 80}, 433--447 (1981) [Sov. Phys. JETP
{\bf 53,} No.~2, 221--227 (1981)]}}

\author{V.G. Nosov$^{\dagger}$ and A.M. Kamchatnov$^{\ddagger}$\\
$^{\dagger}${\small\it Russian Research Center Kurchatov Institute, pl. Kurchatova 1,
Moscow, 123182 Russia}\\
$^{\ddagger}${\small\it Institute of Spectroscopy, Russian Academy of Sciences,
Troitsk, Moscow Region, 142190 Russia}
}

\maketitle

\begin{abstract}
The strictly reversible, thermodynamically equilibrium nature of the free rotation
of a body makes it possible to obtain a number of bounds on the rotational
characteristics within individual rotational bands of nonspherical nuclei.
As a result, the bounds between which the possible values of the critical spin $J_c$
lie can be expressed exclusively in terms of a restricted number of the experimentally
most accessible data on the lower phase $J<J_c$ for a given nuclide. The bounds are
tested on the ground-state rotational bands (yrast lines) of even-even nuclei, in
which the corresponding phase transition (backbending) has already been observed
experimentally. For nuclei with pronounced non-sphericity, all the bounds are
invariably confirmed. For the ground-state rotational bands for which the phase
transition point $J =J_c$ has not yet been reached, predictions are made for the
corresponding values of $J_c^{min}$ and, especially, $J_c^{max}$. The specific features
of excited rotational bands, and also the bands of odd nuclei are discussed.
\end{abstract}

\section{Introduction}

Among the possible collective motions in a nucleus, free rotation occupies a unique
position.   Since it is not accompanied by dissipative processes of frictional type,
it is a thermodynamically equilibrium phenomenon.   This means that, generally speaking,
the strictly reversible interaction of the different degrees of freedom of the system
and the rotation does not lead degradation of the latter.   As long as the shape of
the nucleus is characterized by the presence of the dynamical variable $\mathbf{n}$ (the
unit vector along the symmetry axis of the figure), the corresponding isentropic sequence
of levels---the rotational band---will be continued.

However, the interaction with the other degrees of freedom does, despite being
completely reversible, significantly complicate the situation in precisely the
section of the rotational band hitherto most accessible to experimental study.
The point is that with increasing rotational quantum number $J$ there is an increased
tendency for the mechanical angular momenta of the individual quasiparticles to be
aligned along the direction of the total angular momentum vector $\mathbf{J}$.
Ultimately, this phenomenon will be completely analogous to the alignment of the
magnetic moments of fermions in a magnetic field, and then the rotation properties
will become fairly simple (for more details, see Ref.~[1]).   However, at
low or moderate nuclear spins $J\leq J_c$ (the notation is as before [1--2]) there is
generally not even one single-quasiparticle state within the zone of possible
alignment near the Fermi boundary.   Because of this, the tendency to alignment
along $\mathbf{J}$ is not manifested even here in a pure form.   Coupling of the
nucleons to the axis $\mathbf{n}$ of the nucleus begins to compete with it, and with a
sufficient decrease in the spin $J$ the symmetry of the rotational state is lowered.

Let us consider briefly how these circumstances are reflected formally in the
quantum-mechanical description of rotational states.   It is well known that in the
case of adiabatically slow rotation the total wave function of a nonspherical nucleus
can be represented in the form of the product \footnote{We ignore the circumstance that
$J=\mathbf{J}\cdot\mathbf{n}$ is actually a pseudo-scalar, so that, strictly speaking,
analogous term, corresponding to the value $K=-K_0$, should be added to the
right-hand side of Eq.~(1). In the special and in practice most common and important
case $K_0=0$ of even-even nuclei there is no need for such a term and Eq.~(1) is
valid as it stands.}
\begin{equation}\label{1}
    \Psi_{JM}=\chi_{K_0}(\xi)\left(\frac{2J+1}{4\pi}\right)^{1/2}D_{K_0M}^J(\mathbf{n}),
    \quad J\ll J_c.
\end{equation}
Here, $M=J_z$; $D_{K_0M}^J(\mathbf{n})=D_{K_0M}^J(\varphi,\vartheta,0)$ is a Wigner
function, and $\xi$ is the set of so-called internal variables that the nucleus would
possess if it were not in motion.   But in the general case for arbitrary spins there
is a superposition
\begin{equation}\label{2}
    \Psi_{JM}=\sum_{K=-J}^J\chi_{K}(\xi)\left(\frac{2J+1}{4\pi}\right)^{1/2}D_{KM}^J(\mathbf{n}),
\end{equation}
and the rotational density matrix is expressed as
\begin{equation}\label{3}
\begin{split}
    \rho_{JM}(\mathbf{n},\mathbf{n}')&=\frac{2J+1}{4\pi}\sum_{K=-J}^Jw_K
    D_{K_0M}^J(\mathbf{n})D_{K_0M}^{J*}(\mathbf{n}'),\\
    w_K&=\langle\chi_K^J|\chi_K^J\rangle,
    \end{split}
\end{equation}
where $w_K$ is the probability of the given value of $K$ [on the right-hand side of Eq.~(2),
the internal wave functions $\chi_K^J(\xi)$ are not normalized].

At and above the Curie point $J_c$ we have $\rho_{JM}(\mathbf{n'}, \mathbf{n}) = 1/4\pi=
\mathrm{const}$. Physically, this can be interpreted as the final breaking of
the coupling between a nucleon and the axis of the nucleus, after which the
distribution of the vector $\mathbf{n}$ over the directions in empty space naturally
becomes isotropic, and the quasiparticles tend to be aligned only along $\mathbf{J}$.
Formally, this corresponds to $w_K=(2J+1)^{-1}$ for $J\geq  J_c$, i.e., all $K$ are
equally probable.   For a concrete example, Fig.~1 illustrates the experimental picture
of the phase transition due to the increased symmetry of the rotational state.
In Fig.~1, the phase transition (backbending) can be seen very clearly.
\footnote{It should be noted that in the quantum case in which we are here interested,
the angular velocity $\Omega(J)$ of the rotation is not, in essence, a function
of the state (2).   It is determined by the distance between the two
neighboring rotational levels.   Therefore, the discontinuity of the rotational
velocity in Fig.~1 does not contradict the continuous nature of the phase
transition.}

Since the rotation does not destroy the equilibrium, the thermodynamic treatment
makes it possible to establish a number of bounds for the rotational characteristics.
In this connection, we mention a feature of the present-day experimental situation.
At present, the critical spin $J_c$ has not yet been reached for many nuclei and
their rotational bands, so that the existing data refer exclusively to the lower phase
$J< J_c$.   Do these data predict the position of the phase transition point?
As a rule, this question must be answered in the negative.   However, the thermodynamic
inequalities permit one, by extrapolating the data on the lower phase, to find the
limits between which the true value of $J_c$ lies. In the cases, also fairly numerous,
when the critical value has already been determined experimentally, the reliability
and effectiveness of this procedure can be tested directly.

The overwhelming majority of experimental data of interest in this connection
correspond to the ground-state rotational bands of even-even nuclei (the so-called
yrast lines).   In what follows, we shall have in mind mainly this special case.
The modifications needed in some of the expressions in the more general case will
be indicated separately.

\section{Minimal work for a rotating nucleus and thermodynamic inequalities}

The quantity
\begin{equation}\label{4}
    \widetilde{E}=E(J)-\hbar\Omega_0 J
\end{equation}
is the energy in a coordinate system rotating uniformly with angular velocity
$\Omega_0$ (see, for example, Refs.~3 and 4).   The lowest state $E(J)$ of the
rotational band can always be regarded formally as non-rotating in the sense that
this state minimizes the energy (4) for $\Omega_0 = 0$.   In principle, one can go
over from it to any other state of the band by specifying a corresponding
$\Omega_0=\Omega\neq 0$.   Then the value of $\widetilde{E}$ certainly does not
increase as equilibrium is approached.   Bearing in mind also that
$J=E = \widetilde{E} = 0$ for the original, non-rotating nucleus, we readily obtain
\begin{equation}\label{5}
    E(J)\leq \hbar\Omega J
\end{equation}
for any rotational level.

Applied to rotating bodies, this is entirely equivalent to the notion of the
so-called minimal work [4] which in the given case is $\hbar\Omega J-E$.
In the immediate neighborhood of equilibrium, the energy $\widetilde{E}$ has a
local minimum. This yields the inequality
\begin{equation}\label{6}
    I>0,
\end{equation}
where
\begin{equation}\label{7}
I=\hbar\left(d\Omega/dJ\right)^{-1}=\hbar^2\left(d^2E/dJ^2\right)^{-1}
\end{equation}
is the variable moment of inertia (the Appendix to the preceding
Ref.~1 gives a corresponding simple calculation, though interpreted from a somewhat
different point of view).

So far we have considered inequalities that hold equally in either of the phases.
We now turn to a more definite examination of each of them separately.   We shall
label the lower phase $J \leq J_c-0$ by the index $m$, and the upper, $J>J_c+0$,
by the index $n$.   Where necessary, the additional index $c$ will be used directly
at the Curie point $J=J_c\pm 0$.

First of all, we use some of the results obtained in Refs.~1 and 2 for the upper phase.
It is difficult to visualize clearly its properties near the transition point, at
which the moment of inertia exhibits pole behavior:
\begin{equation}\label{8}
    I\to\frac{j}{J-J_c},\qquad J\to J_c+0
\end{equation}
($j$ is some constant coefficient).   After it has passed through the ``super-rigid-body''
region adjoining the Curie point, the moment of inertia sinks to a minimum, and then
tends to the rigid-body asymptotic form $I=I_0$ from below.   Over the complete upper
phase, the reciprocal value of the moments of inertia cancels out:
\begin{equation}\label{9}
    \int_{J_c}^\infty\left(\frac1{I_0}-\frac1I\right)dJ=0.
\end{equation}

It can be concluded from this that the section from some particular running value $J$
to infinity makes a negative contribution to the integral (9).   As a result, using
Eq.~(7), we readily obtain
\begin{equation}\label{10}
    \frac{\hbar J}{\Omega_n}\geq I_0
\end{equation}
i.e., in the upper phase the ratio of the angular momentum to the angular velocity
exceeds the rigid-body value.   Equality is attained only at the phase transition
point $J= J_c + 0$ itself, and also asymptotically as $J \to\infty$; for more details,
see Ref.~1.

The theoretical investigation of the behavior of the moment of inertia of the lower
phase is made difficult by its low symmetry, and the coupling scheme of the angular
momenta is complicated and itself changes continuously as a function of the spin $J$.
We shall consider this question in detail only for the lowest part of the ground-state
rotational band.   This consists of the levels $J= 0, 2,4,6,\ldots$

The energy $E(J)$ of the levels can be represented as the expectation value of the
Hamiltonian of the nucleus with respect to the true wave function (2), which is a
superposition of different $K$.   But if the expectation value is found in accordance
with the approximate function (1), the result $E^{(0)}(J)$ is an overestimate:
\begin{equation}\label{11}
    E(J)<E^{(0)}.
\end{equation}
We shall assume that the averaging over the trial wave functions (1) is made in two
stages, the first with respect to the internal variables $J.$   Since $K_0 = 0$
for the rotational band in which we are interested, the Wigner functions reduce to
spherical functions, and the new Hamiltonian will act only on them.   Under these
conditions, the scalar operator must be expressed in terms of $\mathbf{J}^2 =J(J+1)$,
and we write the corresponding expansion in the form
\begin{equation}\label{12}
    E^{(0)}(J)=AJ(J+1)-B[J(J+1)]^2+\ldots.
\end{equation}
Here, $A=\hbar^2/2I'$, $I'\equiv I_{J\to0}$ is the adiabatic moment of inertia,
and $B\sim A/J_c^2$.

We expand the true energy $E(J)$ in the usual series in powers of $J$.
In the limit as $J\to 0$ the first and second derivatives of
the energies $E$ and $E^{(0)}$ are equal, since otherwise it would be impossible
to have the adiabatic approximation expressed by the first term on the right-hand side
of Eq.~(12) (the Bohr-Mottelson formula; see, for example, Refs.~5 and 6).   The
difference required by the inequality (11) arises only in the cubic term of the
expansion, and the corresponding coefficient
$$
\left.\frac{d^2E}{dJ^2}\right|_{J\to 0}
$$
has the order of magnitude $\sim\hbar^2/I'J_c\gg B$ (in fact, the expansion is in powers
of the ratio $J/J_c$,; in accordance with the previous Ref.~2, $J_c\sim k_fR\gg1$, where
$k_f$ is the limiting momentum of the Fermi distribution, and $R$ is the radius of the
nucleus).   Thus,
\begin{equation}\label{13}
\left.\frac{d^2E}{dJ^2}\right|_{J\to 0}<0
\end{equation}
and, using Eq. (7), we obtain
\begin{equation}\label{14}
    \left.\frac{dI}{dJ}\right|_{J\to 0}>0,
\end{equation}
i. e., the moment of inertia increases near the base of the rotational band.

The inequality (14) agrees with experiment.   However, numerous experimental data on
the ground-state rotational bands indicate that in them the monotonic growth
\begin{equation}\label{15}
    \frac{dI_m}{dJ}>0
\end{equation}
of the moment of inertia with the spin also holds in the entire lower phase $J<J_c$.

Bearing this in mind, one can draw a number of conclusions about the ratio of the
angular momentum to the angular velocity.   We transform the derivative of this ratio
in accordance with Eq.~(7):
\begin{equation*}
    \begin{split}
    \frac{d}{dJ}\left(\frac{J}{\hbar\Omega_m}\right)&=\frac1{\hbar\Omega_m}
    \left(1-\frac{\hbar^2}I\frac{J}{\hbar\Omega_m}\right)\\
    &=\frac1{\hbar\Omega_m}
    \left(1-\frac{\hbar^2}I\frac{J}{\int_0^J(\hbar^2/I)dJ}\right).
    \end{split}
\end{equation*}
Since
$$
\int_0^J\frac{\hbar^2}IdJ>\frac{\hbar^2}I J,
$$
we have
\begin{equation}\label{16}
    \frac{d}{dJ}\left(\frac{\hbar J}{\Omega_m}\right)>0,
\end{equation}
i.e., the ratio of the angular momentum to the angular velocity also increases.

We now use an inequality obtained earlier in Ref.~2:
\begin{equation}\label{17}
    \Omega_{mc}>\Omega_{nc},
\end{equation}
which determines the sign of the discontinuity of the rotational velocity at
the phase transition point, and the relation
\begin{equation}\label{18}
    \hbar J_c=I_0\Omega_{nc}
\end{equation}
to which Eq.~(9) is essentially equivalent [see also the text following Eq.~(9)].
Then
\begin{equation}\label{19}
    \frac{\hbar J}{\Omega_m}\leq\frac{\hbar J_c}{\Omega_{mc}}<\frac{\hbar J_c}
    {\Omega_{nc}}=I_0.
\end{equation}
We have previously estimated the integral
$$
\int_0^J\frac{\hbar^2}I dJ
$$
from below.   We now obtain an upper bound:
\begin{equation}\label{20}
    \hbar\Omega_m=\int_0^J\frac{\hbar^2}I dJ<\frac{\hbar^2}{I'}J.
\end{equation}
The unified expression of the inequalities (19) and (20) is
\begin{equation}\label{21}
    I'<\frac{\hbar J}{\Omega_m}<I_0.
\end{equation}

In the lower phase, the ratio of the angular momentum to the angular velocity is
less than the rigid-body value but exceeds the adiabatic value $I'$ of this ratio
(in practice, the accuracy of this last assertion is limited by the circumstance
that we have in fact ignored the specifically quantum ``zero-point rotation":
$\hbar\Omega_{J\to 0}=\hbar^2/2I'$). It is also well known that the inequality
$I'<I_0$ agrees with experiment.

\section{Upper and lower bounds for the critical spin $J_c$}

We find first the boundaries of that region on the $(J, E)$ plane within which the
upper phase can in principle exist.   Combining the inequalities (5) and (10), we obtain
\begin{equation}\label{22}
    E_n\leq\frac{\hbar^2J^2}{I_0}.
\end{equation}
This means that the part of the $E(J)$ plot corresponding to the upper phase is
situated entirely to the right of the parabola $\hbar^2J^2/I_0$ [see Fig.~2(a)].
Overall, the curve of the energy of the rotational levels is continuous,
$E_m(J_c) = E_n(J_c)$, since we have a second-order phase transition. Ultimately,
the smallest possible value $J_c^{min}$ of the critical spin is determined by the
transcendental (and in practice empirical) equation
\begin{equation}\label{23}
    E_m(J_c^{min})=\hbar^2[J_c^{min}]^2/I_0,\qquad J_c\geq J_c^{min}.
\end{equation}
The prescription for finding the upper limit $J_c^{max}$ is clear from Eqs.~(10),
(17), and (18); see also Fig.~2(b). In the plane $(J, \hbar\Omega)$ the upper
phase begins on the straight line $\hbar^2J/I_0$ and is situated to the right of it.
The lower phase cannot penetrate to this region, for otherwise the sign of the
discontinuity of the rotational velocity at the phase transition point prescribed
by the inequality (17) would be reversed [the monotonic growth of the moment of
inertia of the lower phase expressed by the inequality (15) does not permit it to
return to the left of the line $\hbar^2J/I_0$; see also Eq.~(7)].   The upshot is
\begin{equation}\label{24}
    \left.\frac{dE_m}{dJ}\right|_{J=J_c^{max}}=\frac{\hbar^2}{I_0}J_c^{max},
    \qquad J_c<J_c^{max}.
\end{equation}
To avoid confusion, we emphasize that the graph of $\hbar\Omega(J)$ is discontinuous
(see also Fig.~1); therefore, $J_c^{max}$ does not coincide with the true $J_c$ but
exceeds it
\footnote{Overall, the above serves sufficiently well for purely practical purposes.
From a deeper and more rigorous point of view, extrapolation of the lower phase
beyond the Curie point is not completely correct mathematically and is somewhat
arbitrary, since the function $E_m(J)$ has a certain singularity here.   However,
we have already noted that at least the first and second derivatives of this
function do not become infinite as $J\to J_c-0$.}.

A typical formulation of the problem is as follows. Suppose that, in contrast to
the example in Fig.~2, there are data on only the lower phase.   We denote $J_f$
the highest value of the spin for which the rotational velocity and the moment of
inertia are known [since calculations in accordance with Eqs.~(7) actually require
the taking of finite differences, the spin $J_F$ of the last of the experimentally
detected levels is in practice usually higher and equal to $J_f+2$].    First, it
is readily seen that as long as $I<I_0$ the graph of $\hbar\Omega(J)$ moves away
from the broken straight line in Fig.~2(b), but when the requirement $I>I_0$ is
satisfied it moves towards this straight line.   This makes it possible to find
an upper bound on the discontinuity $\Delta(\hbar\Omega)=\hbar\Omega_{mc}-
\hbar\Omega_{nc}$ of the angular velocity of rotation of the nucleus:
\begin{equation}\label{25}
    \Delta(\hbar\Omega)_{max}=\hbar\Omega_m(J_c^{min})-\frac{\hbar^2}{I_0}J_c^{min},
    \quad I_m(J_c^{min})>I_0.
\end{equation}
But if $J_f> J_c^{min}$, i. e., the lower limit of the critical spin has already
been passed experimentally, then in Eq.~(25) it is necessary to replace $J_c^{min}$
by $J_f$ (the requirement $I_f>I_0$ remains in force).

We consider the cases in which the finding of the lower or the upper limit requires
a comparatively short extrapolation of the existing data up the band.   Then we can
restrict ourselves to the approximations
\begin{equation}\label{26}
    \begin{split}
    E_m(J)&\approx E_f+\hbar\Omega_f(J-J_f)+\frac{\hbar^2}{2I_f}(J-J_f)^2,\\
    \hbar\Omega_m(J)&\approx\hbar\Omega_f+\frac{\hbar^2}{I_f}(J-J_f).
    \end{split}
\end{equation}
Substitution in Eqs.~(23) and (24) and solution of these equations lead to
\begin{equation}\label{27}
    \begin{split}
    J_c^{min}&=\left\{\frac{\hbar\Omega_f}2-\frac{\hbar^2}{2I_f}J_f+\left[
    \left(\frac{\hbar\Omega_f}2\right)^2-\frac{\hbar^2}{I_0}(\hbar\Omega_fJ_f-E_f)
    -\frac{\hbar^2}{2I_f}\left(E_f-\frac{\hbar^2}{I_0}J_f^2
    \right)\right]^{1/2}\right\}\\
    &\times\left(\frac{\hbar^2}{I_0}-\frac{\hbar^2}{I_f}\right)^{-1},
    \quad\quad\quad J_c^{min}-J_f\lesssim J_f,\\
    J_c^{max}&=\left.\left(\hbar\Omega_f-\frac{\hbar^2}{I_f}J_f\right)\right/\left(
    \frac{\hbar^2}{I_0}-\frac{\hbar^2}{I_f}\right),\quad J_c^{max}-J_f\ll J_f.
    \end{split}
\end{equation}
The criterion of applicability of the first of these is formulated here on the
basis of primarily practical considerations; namely, in the cases of interest, the
quadratic approximation given by the first of Eqs.~(26) usually has a very good
accuracy.   When the conditions of applicability of the expressions are violated,
one must resort to graphical extrapolation.   Naturally, this has its shortcomings.

Finally, we consider a curious application of the boundary curve $\hbar^2J^2/I_0$
on the $(J, E)$ plane.   The less symmetric lower phase is characterized by an order
parameter, whose part is played by the static quadrupole moment $Q$.   Since its
actually realized value must be energetically advantageous, this predetermines the
sign of the discontinuity of the angular velocity in accordance with inequality (17)
(for more details, see the preceding Ref.~2).   One can however also show that there
exists a general restriction of the magnitude of the discontinuity $\Delta\Omega$ as well.

Consider Fig.~2(a).   The plot of $E(J)$ crosses from left to right, the dashed parabola
on which the derivative is $2\hbar^2J_c^{min}/I_0$, and $dE/dJ$ along the band is here
smaller.   Therefore, the inequality (16) enables us to conclude that
$$
\frac{\hbar J_c}{\Omega_{mc}}\geq\frac{\hbar J_c^{min}}{\Omega_m(J_c^{min})}
>\frac{I_0}2.
$$
Expressing now in accordance with (18) the rigid-body moment of inertia in terms of
the upper rotational velocity $\Omega_{nc}$, we find that $\Omega_{mc}<2\Omega_{nc}$,
i.e.,
\begin{equation}\label{28}
    \Delta\Omega<\Omega_{nc}.
\end{equation}
Thus, the magnitude of the abrupt decrease in the rotational velocity at the phase
transition must not exceed half its original value $\Omega_{mc}$.

\section{Comparison with experiment}

For several years, the compilation by Sayer et al [7] served as the prime source of
information on the ground-state rotational bands of individual nuclei. However,
these data are now partly obsolete and many new data have been published.   Therefore,
we have also used original papers [8--16].  References to some other sources of
experimental data that we have used can be found in a later compilation of Lieder and
Ryde [17].

The results for the ground-state rotational bands in which the phase transition has
already been found are summarized in Table I.   The limits $J_c^{min}$ and $J_c^{max}$
were found in accordance with the scheme illustrated in Fig.~2 (see also the text).
When the conditions were more favorable for application of the second of the expressions
(27), the calculated value of the upper limit is given with one decimal.   In the
remaining cases, $J_c^{max}$ was found by graphical extrapolation. Values $J_c^{min}\leq8$
are not given, since they are certainly of no interest. The true critical spin $J_c$ is
determined basically in accordance with Eq.~(18). Sometimes, when suitable data on the
upper phase are available, its value can be found more accurately or confirmed by means
of the relation
\begin{equation}\label{29}
    \hbar\Omega\approx \frac{\hbar^2}{I_0}J_c+\frac{\hbar^2}{2j}(J-J_c)^2,
\end{equation}
which is obtained from Eqs.  (8) and (18) by integrating
the first of them.   Where possible, the result of extrapolation from the upper phase
in accordance with the limiting law (8) is given for comparison.   The details of this
procedure are explained in the previous Ref.~2; we recall here only that it is free of
the necessity of specifying a definite radius of the nucleus (concerning this question,
see also below).

\begin{table}
\begin{tabular}{|c|c|c|c|c|c|c|c|c|c|c|c|c|}
\multicolumn{13}{c}{\bf Table I}\\
\hline
Nucleus & $J_c^{min}$ & $J_c^{max}$  & $J_c$ & $J_c^{extr}$ & $J_f$ &
Nucleus & $J_c^{min}$ & $J_c^{max}$  & $J_c$ & $J_c^{extr}$ & $J_f$ & $r_0$\\
\hline
$^{126}_{56}$Ba$_{70}$ & 11.5 & 20 & 12.7 & -- & -- & $^{164}_{70}
$Yb$_{94}$ & 10.1 & 20 & 13.8 & 13.0 & -- & 1.1\\
$^{128}_{58}$Ce$_{70}$ & 9.3 & 15 & 11.0 & -- & -- &
$^{166}_{70}$Yb$_{96}$ & -- & 19 & 14.4 & 15.6 & -- & 1.1\\
$^{130}_{58}$Ce$_{72}$ & 10.9 & $\sim30$ & 10.6 & 11.1 & -- &
$^{168}_{70}$Yb$_{98}$ & -- & 16.2 & 16.2 & -- & 14 & 1.05\\
$^{132}_{58}$Ce$_{74}$ & 12.4& 17 & 11.2 & -- & -- &
$^{170}_{70}$Yb$_{100}$ & -- & 17 & 15.4 & -- & -- & 1.0\\
$^{134}_{58}$Ce$_{76}$ & $\approx14$ & -- & 10.7 & -- & -- &
$^{168}_{72}$Hf$_{96}$ & 10.8 & 23 & 14.4 & -- & -- & 1.1\\
$^{154}_{64}$Gd$_{90}$ & -- & 17.1 & 16.0 & -- & 14 &
$^{170}_{72}$Hf$_{98}$ & -- & 16.7 & 16.8 & -- & 16 & 1.05\\
$^{154}_{66}$Dy$_{88}$ & 12.6 & 17 & 15.0 & -- & -- &
$^{172}_{72}$Hf$_{100}$ & -- & 16.6 & 16.2 & -- & 14 & 1.0\\
$^{156}_{66}$Dy$_{90}$ & -- & 17 & 15.5 & -- & -- &
$^{170}_{74}$W$_{96}$ & 12.5 & 23 & 13.1 & 13.6 & -- & 1.1\\
$^{158}_{66}$Dy$_{92}$ & -- & 15.7 & 15.7 & 14.8 & 14 &
$^{172}_{74}$W$_{98}$ & -- & 14.9 & 14.8 & 14.4 & 12 & 1.0\\
$^{160}_{66}$Dy$_{94}$ & -- & 16.0 & 15.8 & -- & 14 &
$^{174}_{74}$W$_{100}$ & -- & 18 & 15.6 & -- & -- & 1.0\\
$^{156}_{68}$Er$_{88}$ & 14.4 & $\sim30$ & 14.0 & 13.2 & -- &
$^{176}_{74}$W$_{102}$ & -- & 17 & 15.9 & -- & -- & 1.0\\
$^{158}_{68}$Er$_{90}$ & 12.0 & 19 & 13.5 & -- & -- &
$^{180}_{74}$W$_{106}$ & -- & 16 & 15.8 & -- & -- & 1.0\\
$^{160}_{68}$Er$_{92}$ & 9.2 & 21 & 14.6 & -- & -- &
$^{182}_{76}$Os$_{106}$ & 10.4 & 18 & 12.9 & 11.9 & -- & 1.0\\
$^{162}_{68}$Er$_{94}$ & -- & 20 & 15.3 & 15.7 & -- &
$^{184}_{76}$Os$_{108}$ & 12.4 & 22 & 15.1 & 15.1 & -- & 1.0\\
$^{164}_{68}$Er$_{96}$ & -- & 24 & 15.6 & -- & -- &
$^{186}_{76}$Os$_{110}$ & 14.2 & $\sim30$ & 14.8 & -- & -- & 1.0\\
$^{160}_{70}$Yb$_{90}$ & 13.3 & $\sim30$ & 12.5 & -- & -- &
 &  &  &  &  &  & \\
\hline
\end{tabular}
\end{table}

Naturally, the main aim was to test the validity of the inequalities
$J_c^{min}< J_c<J_c^{max}$.   One sees that there are five violations of the
first of these inequalities, observed for the nuclides $^{130}$Ce, $^{132}$Ce,
$^{134}$Ce, $^{156}$Er, and $^{160}$Yb.   Except for $^{134}$Ce, the violations
of the inequality $J_c>J_c^{min}$ do not in themselves appear too appreciable.
However, the matter appears in a somewhat different light if one notes the
following circumstance:  Near the phase transition point, the inequality (5) is
also violated for the listed nuclides.   Although the width of the region of
violation in the band does not exceed three units of angular momentum, the question
is nevertheless of some interest.   Indeed, in deriving the thermodynamic relation
(5) we assumed essentially only that we have a rotating body (concerning the possible
influence of the spin of the lowest level of the band, see below at the end of this
section).

A likely qualitative explanation is as follows:  In practice, we deduce the rate of
rotation from the distance between neighboring levels, taking half of it from the
equality $\hbar\Omega= dE/dJ$.   But the shape of these nuclei in the ground state
gives grounds for certain doubts, and the deformation of these nuclei evidently varies
along the band.   Under these conditions, one cannot rule out abrupt changes of state;
for example, the deformation may increase abruptly by a certain amount.   If the
positions of the two transitions are close to each other or even coincide, then at
the phase transition point there is an anomalously small distance between the levels,
because the abrupt change in the structure or shape of the nucleus must be energetically
advantageous.   Directly at the point of a first-order phase transition (although in
the given case it is in fact nearly a second-order phase transition with an increase
in symmetry; see also the Introduction), this anomalously small interval does not
correspond to the rotational velocity of the nucleus. By making such an identification,
we significantly reduce the right-hand side of the inequality (5).   After the transition
point has been passed, the intervals between the levels again give the rotational
velocity, and the inequality (5) is again well satisfied.   The fact that our theory
is not always fully adequate for nuclei that may still be spherical in the ground state
has already been noted \footnote{For such nuclides, the previously noted [1] so-called
second backbending is rather characteristic.   At the present time, it has been
found in two nuclei:   $^{158}$Er and $^{160}$Yb.   The reasons for this phenomenon
are not entirely clear; it takes place entirely to the right of and at a depth
$\sim100$~keV below the straight line $\hbar^2J/I_0$.   After this, the graph
$\hbar\Omega(J)$ of the rotational velocity must tend asymptotically to the rigid-body
line $\hbar^2J/I_0$.   However, nuclear spins permitting this tendency to be followed
experimentally have not yet been obtained.}.

In contrast, for nuclei of pronounced nonspherical shape the entire picture of a
second-order phase transition is completely confirmed, and none of the inequalities
given in Sections 2 and 3 is violated.   This also holds for the inequality
$\Delta(\hbar\Omega)<\Delta(\hbar\Omega)_{max}$ [see Eq.~(25) and the text].
The data on the discontinuities of the rotational velocity, whose actual values
are very different (some vanishingly small), are not given in the tables.

Initially, in the calculation of the rigid-body moment of inertia we used the
previously recommended [1-2] value
\begin{equation}\label{30}
    r_0=1.1\cdot 10^{-13}\,\mathrm{cm}
\end{equation}
of the parameter in the well-known expression for the radius of the nucleus.
However, comparison with experiment revealed unexpectedly that for the isotopes of
ytterbium, hafnium, tungsten, and osmium ($Z = 70-76$) at neutron numbers $N>98$
the radius is different.   This can be seen particularly clearly with the
ground-state rotational band of $^{184}_{76}$Os$_{106}$ as the example.
In accordance with Fig.~3, the previous value $r_0 = 1.1\cdot 10^{-13}$~cm is
unsuitable for describing the properties of the upper phase.   Therefore, in
this range of nuclides we gave preference to a smaller radius and used the
working formula \footnote{We have in mind the usual formula $I_0= 2MR^2/5$, where $M$
is the mass of the nucleus.   In the upper phase, isotropy of the distribution
of the vector $\mathbf{n}$ over its spatial orientations corresponds to equal
probability of all directions of the vector $\mathbf{\Omega}\|\mathbf{J}$
with respect to the figure of the nucleus. Under these conditions, the corrections
to the rigid-body moment of inertia that depend on the deformations a could contain
only invariant combination of them, i.e., would actually enter through $\alpha^2$.
We shall throughout ignore the deformation corrections to the rigid-body moment of
inertia, whose relative magnitude is $\sim\alpha^2\ll 1$.}
\begin{equation}\label{31}
    \frac{\hbar^2}{I_0}=\frac{104300}{A^{5/3}}\, [\mathrm{keV}]\quad
    (r_0=1.0\cdot 10^{-13}\,\mathrm{cm})
\end{equation}
instead of formula (15) of Ref.~1.

We do not know the reasons for the decrease in the nuclear radius when $N\geq98$.
Note also that the isotopes of ytterbium $^{168}$Yb and hafnium $^{170}$Hf with
$N=98$ have an intermediate value of the radius $r_0$.

We now consider the ground-state rotational bands in which a phase transition has not
yet been detected experimentally.   The scheme for determining the limits is
illustrated in Fig.~4, and the results are given in Table II.   In the second column,
we give the spin of the last of the experimentally found levels of the band.

With regard to the predictions contained in Table II, we should like to make one remark.
In cases such as $^{152}$Gd, $^{156}$Gd, $^{178}$W, $^{176}$Os, and $^{238}$U, the
experiments have very nearly reached the phase transition point. However, as can be
seen from Eqs.~(24) and (25) and the text, it is precisely in the case of small
$J_c^{max}-J_c$ that the discontinuity $\Delta(\hbar\Omega)$ is negligible.
If the discontinuity of the rotational velocity is not discerned in experiment,
the phase transition is by no means so striking as in Figs.~1, 2(b), and 3.
Transition to the upper phase must be gauged from the fulfillment of the inequality
(10) at the achieved interval between the levels. But if we are not satisfied with
this and wish to deduce the arrival in the upper phase also from the course of the
moment of inertia (which is not related to concrete assumptions about the radius of
the nucleus), then one or two more energy intervals are required.

\begin{table}
\begin{tabular}{|c|c|c|c|c|c|c|c|c|}
\multicolumn{9}{c}{\bf Table II}\\
\hline
Nucleus & $J_F$ & $J_c^{min}$ & $J_c^{max}$  &
Nucleus & $J_F$ & $J_c^{min}$ & $J_c^{max}$  & $r_0$\\
\hline
$^{150}_{58}$Ce$_{92}$ & 8 & -- & 15 & $^{174}_{70}$Yb$_{104}$ & 20 & -- & 22.9 &  1.0\\
$^{134}_{60}$Nd$_{74}$ & 8 & 12 & 18 & $^{176}_{70}$Yb$_{106}$ & 18 & -- & 20.9 &  1.0\\
$^{136}_{60}$Nd$_{76}$ & 8 & $\approx126$ & $\sim30$ & $^{166}_{72}$YHf$_{94}$ & 14 & -- & 19 &  1.1\\
$^{150}_{60}$Nd$_{90}$ & 8 & -- & $\sim20$ & $^{174}_{72}$YHf$_{102}$ & 14 & -- & 17.3 &  1.0\\
$^{150}_{62}$Sm$_{88}$ & 12 & 12.7 & 16 & $^{176}_{72}$Hf$_{104}$ & 14 & -- & 21 &  1.0\\
$^{152}_{62}$Sm$_{90}$ & 10 & -- & 17 & $^{178}_{74}$W$_{104}$ & 16 & -- & 17.5 &  1.0\\
$^{154}_{62}$Sm$_{92}$ & 10 & -- & 17 & $^{174}_{76}$Os$_{98}$ & 10 & -- & 13 &  1.0\\
$^{152}_{64}$Gd$_{88}$ & 16 & 12.2 & 16.9 & $^{176}_{76}$Os$_{100}$ & 16 & -- & 18 &  1.0\\
$^{156}_{64}$Gd$_{92}$ & 16 & -- & 15.6 & $^{178}_{76}$Os$_{102}$ & 16 & 8.1 & 20.2 &  1.0\\
$^{162}_{66}$Dy$_{96}$ & 18 & -- & 23 & $^{180}_{76}$Os$_{104}$ & 14 & 9.5 & 17.0 &  1.0\\
$^{164}_{66}$Dy$_{98}$ & 12 & -- & 21 & $^{232}_{90}$Os$_{142}$ & 18 & -- & 24 &  1.1\\
$^{166}_{68}$Er$_{98}$ & 16 & -- & 19.1 & $^{2328}_{92}$U$_{146}$ & 24 & -- & 26.7 &  1.1\\
$^{162}_{70}$Yb$_{92}$ & 12 & 12.8 & 20 &  &  &  &  &  \\
\hline
\end{tabular}
\end{table}

Besides the ground-state rotational bands, the energy
spectra of nonspherical nuclei also contain, of course, other bands with
$J= J',$ $J' + 2,$ $J' + 4,$ $J' + 6$\ldots; in the general case $J'\neq 0$.
Then the energy $\widetilde{E}$ determined by Eq.~(4) (in equilibrium
$\Omega=\Omega_0$) is also nonzero for a ``non-rotating" lowest state $J= J'$
of the band. Finally, the generalization of the inequality (5) takes the form
\begin{equation}\label{32}
    E-E'\leq\hbar\Omega(J-J'),
\end{equation}
where $E'$ is the excitation energy of the level $J=J'$. Accordingly, instead
of (23) we arrive at the equation
\begin{equation}\label{33}
    E_m(J_c^{min})-E'=\hbar^2J_c^{min}(J_c^{min}-J')/I_0.
\end{equation}

After substitution of the quadratic approximation given by the first of equations
(26), the corresponding solution appears somewhat cumbersome:
\begin{equation}\label{34}
    \begin{split}
    J_c^{min}=\left(\frac{\hbar^2}{I_0}-\frac{\hbar^2}{2I_f}\right)^{-1}\Bigg\{
    \frac12\left(\hbar\Omega_f+\frac{\hbar^2}{I_0}J'-\frac{\hbar^2}{I_f}J_f\right)+
    \Bigg[\frac14\left(\hbar\Omega_f-\frac{\hbar^2}{I_0}J'\right)^2\\
    -\frac{\hbar^2}{I_0}[\hbar\Omega_f(J_f-J')-(E_f-E')]-\frac{\hbar^2}{2I_f}
    \left[(E_f-E')-\frac{\hbar^2}{I_0}J_f(J_f-J')\right]\Bigg]^{1/2}\Bigg\},\\
    J_c^{min}-J_f\lesssim J_f.
    \end{split}
\end{equation}
When the applicability of this formula does not inspire particular confidence,
Eq.~(33) must be solved by graphical extrapolation (or interpolation if the value
$J_c = J_c^{min}$ has already been passed through experimentally) of the data on the
lower phase.

With regard to Eq.~(24), and also the second of the approximate expressions (27),
they remain valid. Quite generally, it should be noted that the basic properties of
the upper phase and the form of the corresponding relations and inequalities do not
depend on the spin $J'$ of the lowest level.   It has, for example, the very
characteristic asymptotic tendency to the simple rigid-body law $\hbar J=I_0\Omega$ of
proportionality between the angular momentum and the angular velocity as $J-J_c\to\infty$.
This qualitative feature does not depend on the specific choice of the band.   Nor does
it depend on the absence or presence of an odd nucleon or on the value of the individual
angular momentum that one is inclined to ascribe to it in a particular model of slow
rotation (in making this last comment, we have in mind odd nuclei).

As an example, we consider some data obtained in the experimental study of Ref.~12.
We are here concerned with the position of the energy levels of the nuclide $^{164}$Er,
which evidently correspond in the region of adiabaticity of the rotation to $\gamma$
vibrations of the nucleus (quadrupole shape vibrations corresponding to departures
from axial symmetry of the figure).   The evaluation results are as follows:
$$
\begin{array}{l}
{\rm Band}\quad 2^+,\,4^+,\,6^+,\ldots J_F=18;\quad J_c^{min}=8.8;\quad J_c^{max}\gtrsim 20;
\quad J_c=15.2.\\
{\rm Band}\quad 3^+,\,5^+,\,7^+,\ldots J_F=21;\quad J_c^{min}=10.9;\quad J_c^{max}=18;
\quad J_c=14.0.\\
\end{array}
$$
The pole dependence of the moment of inertia of the upper phase of this band, which is
fairly clearly pronounced in accordance with the limiting law (8), also permits the
estimates $J_c^{extr}= 14.8$ and $j/I_0 = 2.1$.

It should however be said that the non-monotonicity of the moment of inertia of the
lower phase sometimes observed in the excited rotational bands of even-even nuclei can
introduce some uncertainty in the estimate of $J_c^{max}$.

For odd nuclei, $J' \neq 0$ in any band.   For such bands, Eq.~(33) is suitable as is,
in favorable cases, its approximate solution (34).   We have not analyzed here
examples corresponding to odd nuclei.

\section{Discussion}

The theory of the phenomenon in which we are interested, which is based on the notion
of non-conservation of the quantum number $K$, would be verified best on the basis of
the static quadrupole moments and the intensities of quadrupole transitions between
neighboring rotational levels.   However, in the high-spin states in which we are
interested, the static quadrupole moments have not been measured at all.   As yet,
data on $E2$ transitions are fragmentary and their accuracy leaves something to be
desired.   In addition, an appreciable fraction of these data corresponds to the
region $K\approx0$ of adiabatic slowness of the rotation, where, naturally, they do
not appear to contradict the well-known standard expressions.   It is precisely
where the predictions of the theory are particularly unambiguous (in the upper phase)
that the experimental points can be literally counted and, as a rule, have large errors.

The true intensity of an $E2$ transition is usually divided by its purely adiabatic
value calculated in accordance with the model wave function (1) ($K_0 = 0$).
In accordance with the previous Ref.~2, $F_{theor} = 0.48$ at $J= 14$ for this
relative intensity.   We give the recently published experimental result [18], which
has a more or less acceptable accuracy:  In the case of $^{126}$Ba, the transition
$14^+ \to 12^+$ was found to have intensity $F = 0.53_{-0.08}^{+0.26}$ (the result is
taken from the figure in Ref.~18).   At the present state of the art, there is
apparently no contradiction between theory and experiment \footnote{Speaking of the complete
set of accumulated experimental data, one can establish that the value $F= 1$,
which corresponds to $K= {\rm const}= 0$, is fairly convincingly refuted in
the general case by the already existing data.}.

But if one is not satisfied with such accuracy, it is necessary to resort to a
more indirect verification of the theory, based on the arrangement of the rotational
levels themselves, to which the present work is in fact devoted.

\bigskip

\bigskip

\centerline{\bf Figures captions}

\bigskip

Fig.~1.   Phase transition (backbending) in the ground-state rotational band of $^{170}$W.

\bigskip
Fig.~2. Lower (a) and upper (b) limits for the possible values of the critical spin
$J_c$ (ground-state rotational band of $^{164}_{70}$Yb$_{94}$).

\bigskip
Fig. 3.   Change in the slope of the rigid-body line $\hbar^2J/I_0$ for example of
rotation of the nucleus $^{184}_{76}$Os$_{106}$:
1)  $r_0= 1.0\cdot 10^{-13}\,\mathrm{cm}$, 2) $r_0= 1.1\cdot 10^{-13}\,\mathrm{cm}$.

\bigskip
Fig.~4.  Scheme for determining $J_c^{min}$ (a) and $J_c^{max}$ (b) by extrapolation
of the data of the lower phase (the case of $^{162}_{70}$Yb$_{92}$).


\begin{thebibliography}{99}

\bibitem{1} V.G. Nosov and A.M. Kamchatnov, Zh. Eksp. Teor. Fiz.
{\bf 76,} 1506 (1979) [Sov. Phys. JETP {\bf 49,} 765 (1979)] eprint nucl-th/0405001.

\bibitem{2} V.G. Nosov and A.M. Kamchatnov,  Zh. Eksp. Teor. Fiz.
{\bf 73,} 785 (1977) [Sov. Phys. JETP {\bf 46,} 411 (1977)] eprint nucl-th/0403083.

\bibitem{3} L.D. Landau and E.M. Lifshitz, {\it Mekhanika,} Nauka,  Moscow
(1973);  English translation: {\it Mechanics}, Oxford (1976).

\bibitem{4} L.D. Landau and E.M. Lifshitz, {\it Statisticheskaya Fizika,}
Nauka, Moscow (1976),  Part I; English translation:  {\it Statistical Physics,}
2 vols., 3rd ed., Pergamon Press,  Oxford (1980).

\bibitem{5} L.D. Landau and E.M. Lifshitz, {\it Kvantovaya mekhanika,}
Nauka, Moscow (1979); English translation:  {\it Quantum
Mechanics,} Pergamon Press, Oxford (1973).

\bibitem{6} A. Bohr and B.R. Mottelson, {\it Nuclear Structure,} Vol. 2,
New York (1969) (Russian translation published by Mir,
Moscow (1977)).

\bibitem{7} R.O. Sayer, J.S. Smith, and W.I. Milner, At. Data Nucl.
Data Tables {\bf 15,} 85 (1975).

\bibitem{8} C. Flaum, D. Cline, A.W. Sunyar, O.C. Kistner, Y.K. Lee,
and J.S. Kim, Nucl. Phys. A {\bf 264,} 291 (1976).

\bibitem{9} D. Ward, H.R. Andrews, O. Hausser, Y. El-Masri, M.M.
Aleonard, I. Yang-Lee, R.M. Diamond, F.S. Stephens, and
P.A. Butler, Nucl. Phys. A {\bf 332,} 433 (1979).

\bibitem{10} M.W. Guidry, I.Y. Lee, N.R. Johnson, P.A. Butler,  D.
Cline, P. Colombani, R.M. Diamond, and F.S. Stephens,
Phys. Rev. C {\bf 20,} 1814 (1979).

\bibitem{11} I.Y. Lee, M.M. Aleonard, M.A. Deleplanque, Y. El-Masri,
J.O. Newton, R.S. Simon, R.M. Diamond, and F.S. Stephens, Phys. Rev. Lett.
{\bf 38,} 1454 (1977).

\bibitem{12} N.R. Johnson,  D. Cline, S.W. Yates, F. Stephens, L.L.
Riedinger, and R.M. Ronninger, Phys. Rev. Lett. {\bf 40,} 151 (1978).

\bibitem{13} F.A. Beck, E. Bozek, T. Byrski, C. Gehringer, J.C.
Merdinger, Y. Schutz, J. Styczen, and J.P. Vivien, Phys.
Rev. Lett. {\bf 42,} 493 (1979).

\bibitem{14} G.D. Dracoulis,  P.M. Walker, and A. Johnston, J. Phys.
G {\bf 4,} 713 (1978).

\bibitem{15} C.L. Dors,  F.M. Bernthal, T. . Khoo,  C.H. King, J.
Borggreen, and G. Sletten, Nucl. Phys. A {\bf 314,} 61 (1979).

\bibitem{16} F.M. Bernthal, C.L. Dors, T.L. Khoo, and R.A. Warner,
Phys. Lett. B {\bf 64,} 147 (1976).

\bibitem{17} R.M. Lieder and H. Ryde, Adv. Nucl. Phys. {\bf 10,} 1 (1978).

\bibitem{18} G. Seller-Clark,  D. Husar, R. Novotny, H. Graf, and D.
Pelte,  Phys. Lett. B {\bf 80,}, 345 (1979).

\end{thebibliography}
\end{document}